\documentclass[12pt,preprint]{emulateapj}
\usepackage{mathptmx,psfig,pstricks}

\slugcomment{To Appear in The Astrophysical Journal} 



\newcommand{\psr}{PSR~J0205+6449}
\newcommand{\snr}{3C~58}

\newcommand{\chandra}{{\em Chandra}}

\begin{document}

\title{
New Constraints on the Structure and Evolution of the Pulsar Wind Nebula 3C 58
}

\author{Patrick Slane\altaffilmark{1}, David J. Helfand\altaffilmark{2},
Eric van der Swaluw \altaffilmark{3}, and Stephen S. Murray\altaffilmark{1}}

\altaffiltext{1}{Harvard-Smithsonian Center for Astrophysics,
    60 Garden Street, Cambridge, MA 02138}
\altaffiltext{2}{Columbia Astrophysics Laboratory, Columbia University,
550 W. 120th St., New York, NY 10027}
\altaffiltext{3}{FOM-Institute for Plasma Physics Rijnhuizen,
PO Box 1207, 3430 BE Nieuwegein, The Netherlands.}

\begin{abstract}
We present an investigation of the spectral and spatial structure of the X-ray 
emission from \snr\ based on a 350~ks observation with the {\it Chandra X-ray
Observatory}. This deep image, obtained as part of the \chandra\ Large Project
program, reveals new information on nearly all spatial scales in the pulsar
wind nebula (PWN). On the smallest scales we derive an improved limit 
of $T < 1.02 \times 10^6$~K for blackbody emission from the entire surface 
of the central neutron
star (NS), confirming the need for rapid, nonstandard cooling in the stellar
interior. Furthermore, we show that the data are consistent with emission
from a light element atmosphere with a similar temperature. Surrounding 
the NS, a toroidal structure with a jet is resolved, consistent with earlier 
measurements and indicative of an east-west orientation for the projected 
rotation axis of the pulsar. A complex of loop-like X-ray filaments fills 
the nebula interior, and corresponds well with structures seen in the radio 
band. Several of the structures coincide with optical filaments as well. The 
emission from the interior of the PWN, including the pulsar, jet, and 
filaments, is primarily nonthermal in nature. The power law index steepens 
with radius, but appears to also show small azimuthal variations. The 
outermost regions of the nebula require a thermal emission component, 
confirming the presence of an ejecta-rich swept up shell.

\end{abstract}


\section{Introduction}

The Crab pulsar, the product of a core-collapse supernova in 1054 CE,
broadcasts
its presence through a surrounding pulsar wind nebula (PWN) that is one of the
most luminous radio and X-ray sources in the Galaxy. Its probable
Medieval sibling, thought to have been
born in the supernova of 1181 CE, is more modest. Its radio PWN, 3C58, is less
luminous than the Crab by an order of magnitude, and its X-ray nebula is 2000 
times weaker. In part, these differences are due to the birth properties of the
two pulsars: the Crab's initial spin period of $\sim 19$~msec means that nearly
$5 \times 10^{49}$~ergs has been dumped into its PWN in the last 950 yrs;
it maintains a current spindown power of $4.6 \times 10^{38}$ erg s$^{-1}$. 
In contrast, PSR J0205+6449 in 3C58 requires a birth period of $\sim 60$~ms
to explain its current period of 65~ms (Murray et al. 2002),
and provides only $2.7 \times 10^{37}$~erg s$^{-1}$ to its PWN today -- a
factor of nearly 20 smaller than the Crab, albeit the second highest spin-down
power known for a pulsar in the Galaxy.

Differences in the energetics of the powering neutron stars may well not be the
only characteristic that distinguishes 3C58 from the Crab Nebula. Despite
apparently being younger, 3C58 is considerably larger and yet 
exhibits lower expansion
velocities among its optical filaments. Optical observations
(Fesen 1983) reveal knots and filaments which show velocities
$v_r \approx +900$ to $-700 {\rm\ km\ s}^{-1}$, and radio expansion
measurements (Bietenholz, Kassim, \& Weiler 2001) yield a similar result.
Thus, the current expansion rate is considerably less than the undecelerated
velocity $v \approx 3.9 \times 10^3 {\rm\ km\ s}^{-1}$ required to yield the
current size of 3C~58 if the age is only 820 yr. This has led some authors
to suggest that 3C58 is not the remnant of SN1181. However, a recent review of
Chinese and Japanese manuscripts (Stephensen \& Green 2002) lends strong
support
to the association; of the known SNRs within a $10^\circ \times 10^\circ$
region
that best corresponds to the supernova position inferred from these early
records, only 3C~58 has properties consistent with a very young object. The
presence of the second most energetic pulsar known further reinforces the
connection. 

\begin{figure*}[tb]
\pspicture(0,13)(18.5,21)

\rput[tl]{0}(0.0,21.0){\psfig{file=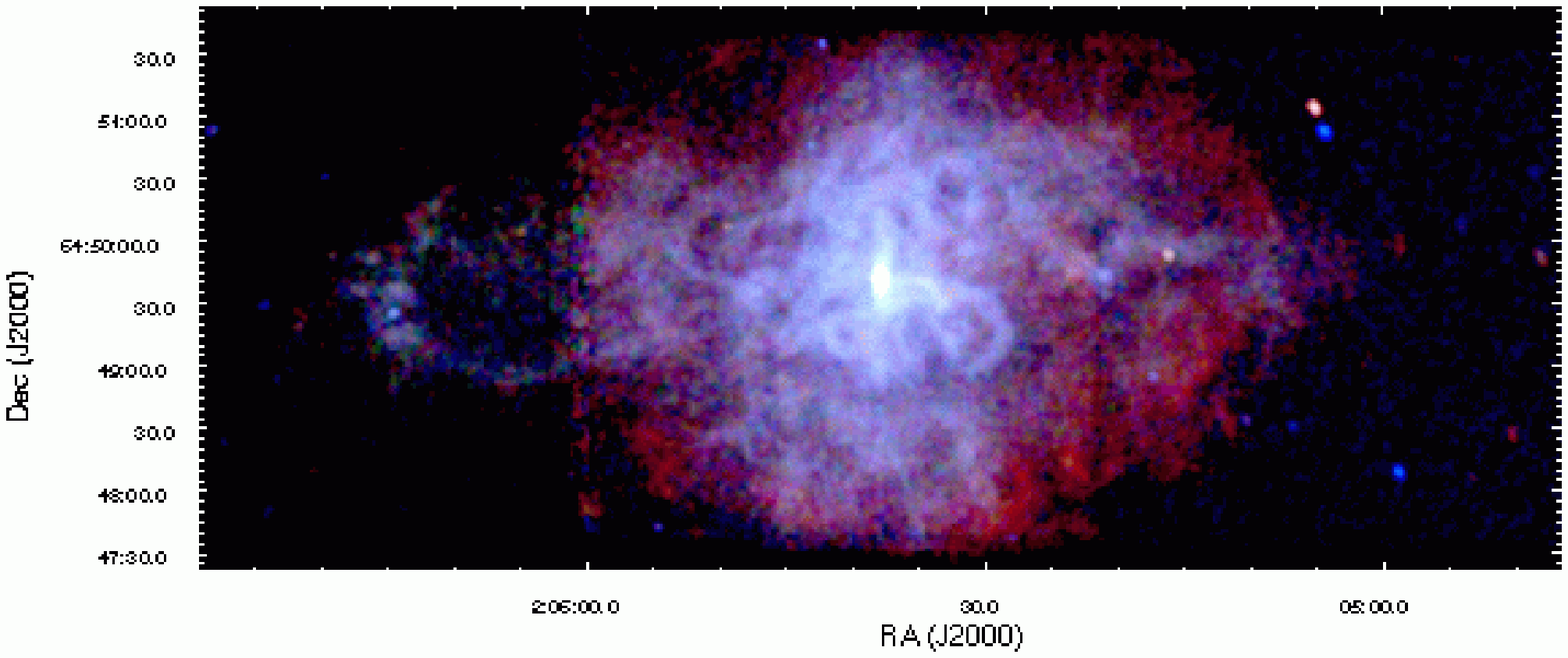,width=16.5cm}}

\rput[tl]{0}(0,14.0){

\begin{minipage}{16.25cm}
\small\parindent=3.5mm
{\sc Fig.}~1.---
ACIS image of 3C 58 from co-added images in the energy bands 0.5-1.0~keV (red),
1.0-1.5~keV (green), and 1.5-10 keV (blue). The pulsar is at the center, and
is surrounded by a elongated compact nebula with a curved jet extending to the
west. Complex filamentary loops fill the interior region, and a softening of
the spectrum with radius is evidenced by the red outer regions -- an
effect resulting from both synchrotron aging of the electrons and the
presence of a soft thermal shell.
\end{minipage}
}
\endpspicture
\end{figure*}

X-rays were first detected from 3C58 by the Einstein Observatory (Becker, 
Helfand, and Szymkowiak 1982). These observations
showed the centrally peaked brightness distribution and power law spectrum
expected from a Crab-like remnant, confirming the classification of this source
by Wilson and Weiler (1976). The surface brightness profile of the nebula
suggested the presence of a central point source, but it was twenty years
before the pulsar was finally discovered at X-ray (Murray et al. 2002) and
radio
(Camilo et al. 2002) wavelengths. ROSAT, ASCA, XMM, and Chandra observations
of 3C58 added further details to our picture of the PWN,
resolving the central enhancement (Helfand, Becker and White 1995), measuring 
the steepening of the X-ray spectrum with increasing radius (Torii et al.
2000),
finding evidence for a shell of thermal emission (Bocchino et al. 2001), and 
setting a limit on the neutron star temperature (Slane, Helfand and Murray 
2002). The rich detail hinted at by the image obtained in the latter
publication
inspired us to propose a very deep observation of the nebula. The result of
that observation is presented here.

In section 2, we describe the observations, while section 3 includes details of
our spatial and spectral analysis. We then go on to discuss the large-scale
structure of the nebula ($\S 4.1$), the geometry of the inner nebula ($\S4.2$),
improved limits on the neutron star temperature ($\S 4.3$), the jet-like 
extension emanating from the pulsar ($\S 4.4$), the remarkable filamentary loop
structures revealed in the image ($\S 4.5$), and the thermal shell emission 
($\S 4.6$). Section 5 summarizes our conclusions.

\begin{figure*}[tb]
\pspicture(0,-1)(18.5,21)
\rput[tl]{0}(2.0,21){\psfig{file=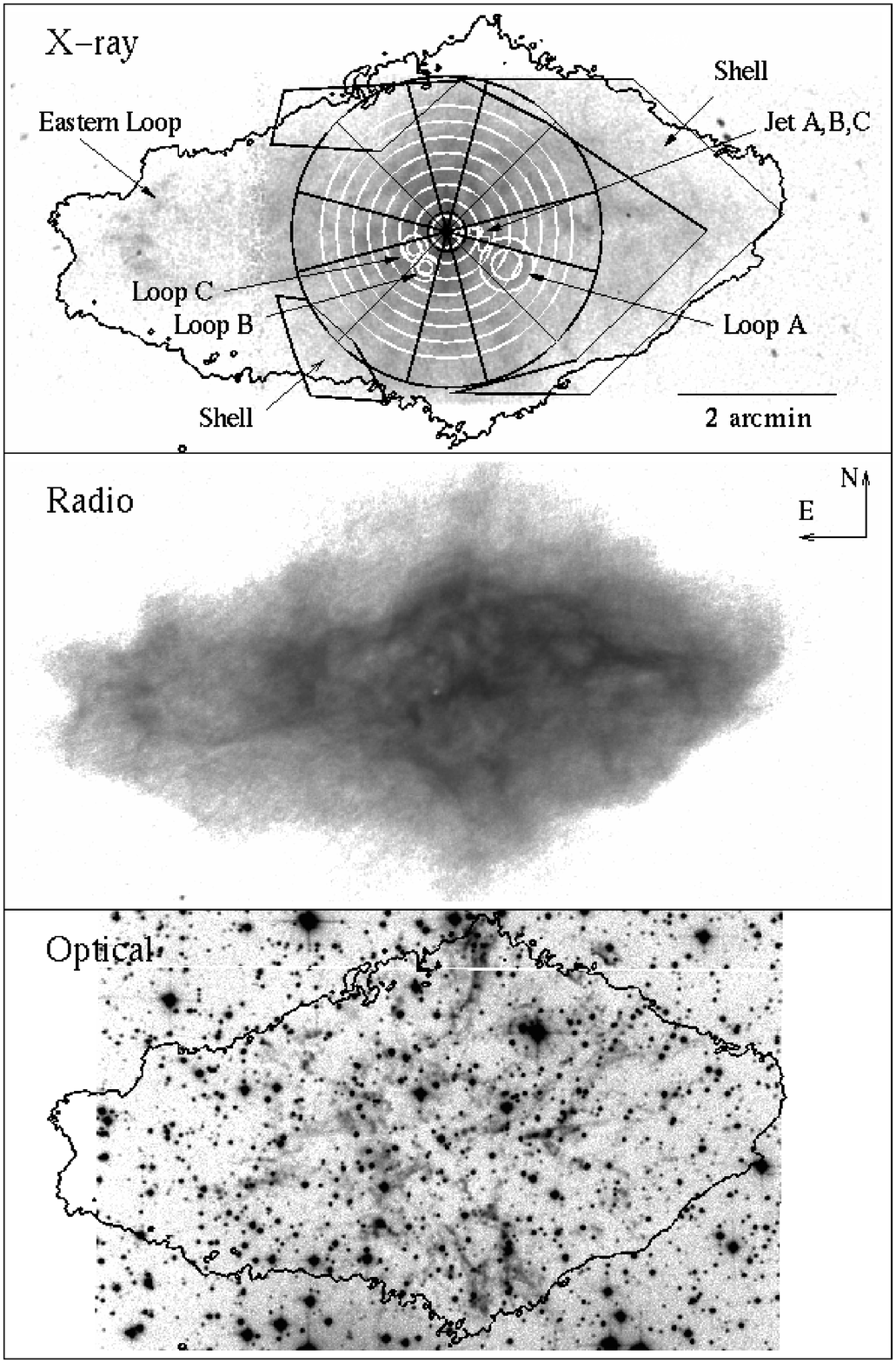,width=13cm}}

\rput[tl]{0}(0,0.0){
\begin{minipage}{16.25cm}
\small\parindent=3.5mm
{\sc Fig.}~2.---
Top: Chandra image of 3C~58 showing spectral extraction regions referred to
in the text. The outermost contour from the radio image is shown for reference.
Middle: VLA image of 3C 58, from Reynolds \& Aller 1988 (kindly provided by
S. Reynolds). Virtually all of the filamentary structures observed in the
X-ray image have counterparts in the radio image. (The ``hole'' observed near
the center of the image is an artifact explained in Reynolds \& Aller 1988).
Bottom:
H$\alpha$ image of 3C~58 showing optical filaments in nebula interior.
The circular filamentary region in the south is also observed in the X-ray
and radio images.
\end{minipage}
}
\endpspicture
\end{figure*}

\section{Observations}

3C~58 was observed for 350~ks with the ACIS detector onboard the \chandra\
Observatory between 22-26 April 2003.
The aimpoint was placed on the S3 chip of the detector, and a
1/2-subarray mode was selected in order to minimize pileup from the compact
central source while still imaging most of the PWN. 
The observation was divided into three segments (ObsIDs 4383, 4382, and
3832 in time order) to accommodate interruptions by the radiation zones in 
the \chandra\ orbit. An error in on-board commanding resulted in the final
segment being carried out with the spacecraft dither turned off for the full
170~ks duration, resulting in no exposure in regions corresponding to gaps 
between the CCD chips or to bad column regions in the CCDs for this segment.

Standard cleaning of the data to remove episodes of high background,
which affected $\sim 15\%$ of the middle segment, resulted 
in a final exposure of 317~ks. The X-ray image, shown in Figure 1 and 
described in detail below, reveals a complex structure consisting of loops, 
elongated features, and broad diffuse emission. 

\section{Analysis}

\subsection{Spatial Analysis}
The exposure-corrected X-ray image of 3C~58 is shown in Figure 1. Here we have 
blocked the raw image by a factor of two, smoothed with a 1-bin 
(0.5~arcsec) Gaussian,
and divided by the exposure map. The upper and lower edges of the image
are set by the boundaries of the subarray readout. Faint vertical lines
in the image are small artifacts related to the lack of dither in one 
observation segment, as described above.
To produce the ``X-ray color'' image shown, color planes were created
for three energy bands: 0.5--1.0~keV (red), 1.0--1.5~keV (green)
and 1.5--12~keV (blue). As we discuss below, the color image reveals
distinct energy-related structure, including an outer shell of soft
emission.

The central region of 3C~58 is dominated by emission from the pulsar
\psr\ (Murray et al. 2002) which produces an 
extended structure, with elongation in the N-S direction, perpendicular to 
the long axis of the main nebula. Slane, Helfand, \& Murray (2002) 
interpret this emission as a tilted ring-like equatorial structure associated 
with the pulsar wind termination shock, where the fast-moving pulsar wind
merges with the much slower expansion of the nebula itself. They also noted the 
presence of a jet-like feature protruding $\sim 30$~arcseconds to the west, 
perpendicular to the inferred equatorial
axis. This feature is particularly evident in Figure 1, and presumably 
corresponds to the rotation axis of the pulsar. 

A striking feature of the X-ray morphology is the presence of numerous 
loop-like structures with radii of $\sim 10-15$~arcsec. The size-scale
of these remarkable features, and their geometric distribution, are
presumably signatures of a complex and tangled nebular magnetic field structure
in the nebula. We discuss their physical characteristics in more detail
in Section 4.

As illustrated in Figure 2, the radio morphology of 3C~58 is quite
similar to that observed in X-rays. While not immediately evident in the 
radio image, a comparison with the \chandra\ image reveals radio structures
that coincide with each of the X-ray loops. Moreover, the
overall extent of the nebula is nearly the same in the two spectral
bands, as is the structure of several other large-scale features such as
the large loop in the eastern portion of the nebula. The nearly identical
morphology is in complete contrast to the Crab Nebula, for example, in which
the X-ray emission is confined to a much smaller region due to the short
synchrotron lifetime of the high energy electrons. For 3C~58, the X-ray
emitting particles persist nearly all the way to the edge of the radio nebula
(Figure 3), indicating that synchrotron losses are not significant up to very 
high electron energies. 

\setcounter{figure}{2}
\begin{figure}[t]
\plotone{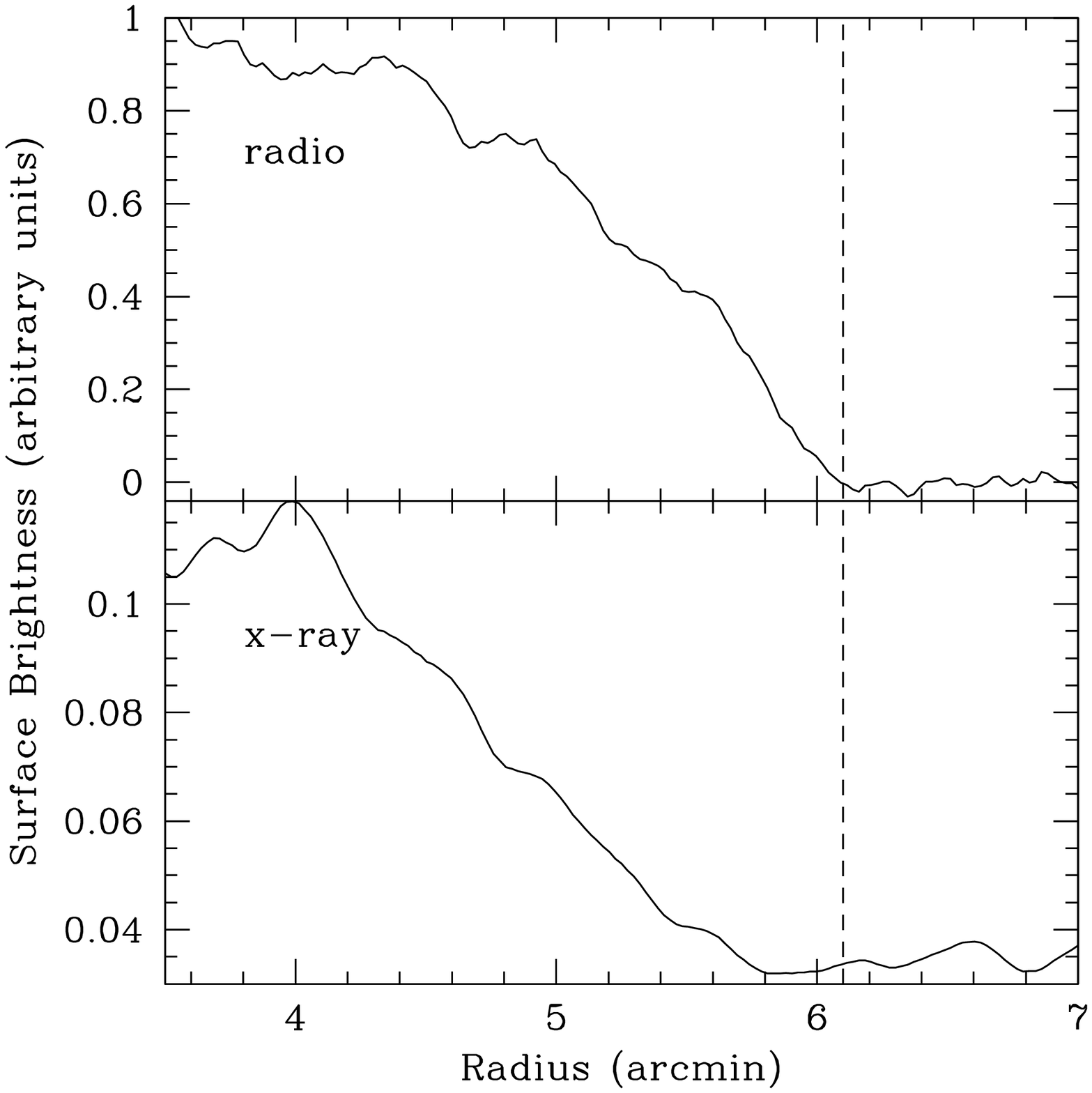}
\caption{
Comparison of radio and X-ray extent of 3C~58. The brightness profile
for a radial slice in the southwest quadrant is shown for the radio (top)
and X-ray (bottom) bands. The dashed line demarcates the edge of the
radio nebula. Although the falloff is gradual, the X-ray emission extends
to within about 20 arcseconds of the radio boundary.
}
\end{figure}

\subsection{Spectral Analysis}

X-ray spectra were extracted from various regions of 3C~58, with corresponding
background spectra taken from a region of the CCD beyond the boundary of the
nebula. Prior to extraction, a time-dependent gain correction was applied
to the data using the {\tt tgain\_corr} software developed by 
A. Vikhlinin. Weighted spectral response and effective area tables for the 
extended emission regions were generated with the {\tt acisspec} routine in
{\tt ciao 3.0.2}. The reduced effective area at low energies caused by 
contaminants that have condensed on the ACIS filter has been accounted for in
the effective area tables.

The emission from 3C~58 is dominated by a power law component, typical of
synchrotron emission. 
However, as described below, a faint thermal component is clearly detected
in the outer regions of the PWN, and also contributes to the interior
regions. Using a two component model consisting of a power law and a thermal
plasma in ionization equilibrium (see Section 4.6), we performed joint fits
to spectra from the annular regions indicated in Figure 2. We obtain a
best-fit column density of $N_H = (4.53 \pm 0.09) \times 10^{21}{\rm\ cm}^
{-2}$. Fitting individual regions separately (both the annuli and also
wedge-shaped regions shown in Figure 2) we do not find any significant
evidence of spatial variations in $N_H.$ For regions in which the count
rate is to low to effectively constrain the absorption, we adopt this value. 
The uncertainties quoted here, and throughout the paper (unless otherwise
noted), refer to 90\% confidence intervals
determined from chi-squared fitting of the data to the specified models.

The spectral index varies throughout 3C~58.
We find, in particular, that the average spectral index increases with radius,
as expected for synchrotron losses as the electrons diffuse away from the
central regions, and consistent with the results based on studies of 3C~58 with 
{\em ASCA} (Torii et al. 2000) and {\em XMM-Newton} (Bocchino et al. 2001).
Such an effect is also seen in G21.5--0.9 (Slane et al. 2000, 
Warwick et al. 2001) and other young PWNe. In 3C~58 the X-ray nebula
extends nearly all the way to the edge of the radio nebula, indicating
that the spectral break falls just below the soft X-ray band; only high
energy electrons are significantly burned-off before reaching the outer
regions of the nebula.
In Figure 4 we plot the radial variation of the power law index in
3C~58, obtained by extracting spectra from annular regions centered on the
pulsar. Here we have used only energies from $2.2 - 8$~keV in the fit in order
to eliminate any contributions from a soft thermal component (see Section 4.6).

\begin{figure}[t]
\plotone{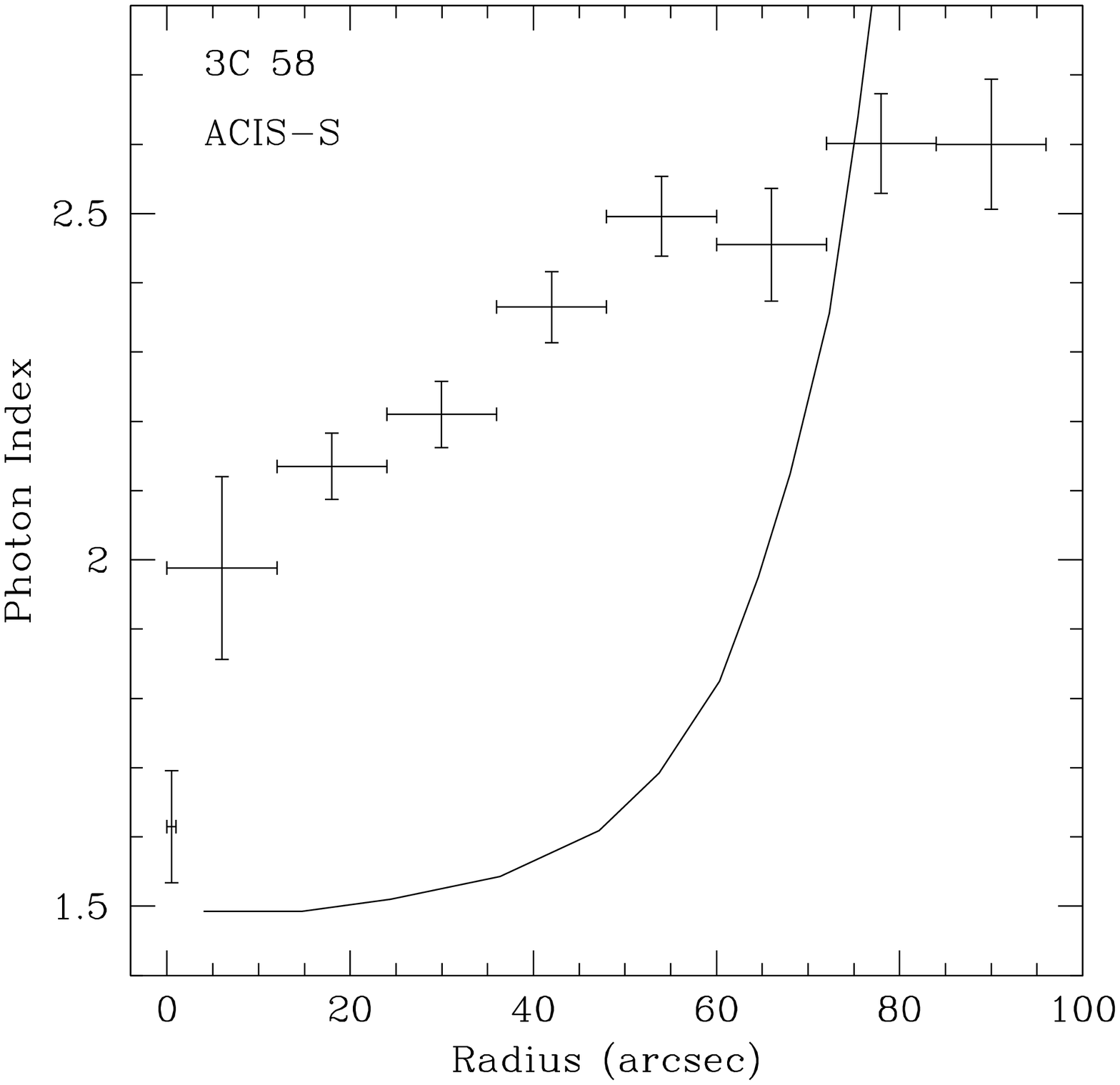}
\caption{
Variation in spectral index with radius in 3C~58. The solid curve is
the predicted variation based on the KC84 model (from Reynolds 2003);
see Section 5.
}
\end{figure}

For the jet-like region protruding from the pulsar, we extracted spectra
from the entire structure as well as three distinct regions extending 
across the feature. Fixing
the column density for each region at the value derived above, we find a
spectral index of $\sim 2.1$ with no significant evidence for 
variations along the jet (see Table 1).

As noted above, one of the remarkable features revealed by the deep 
ACIS observation of 3C~58 is the presence of loop-like filaments throughout
the nebula. We have investigated the spectra from several 
individual filaments, and have compared these with the emission in the 
``void'' regions interior to the loops to search for evidence of variations 
that might be expected if the synchrotron-emitting particles age as they 
diffuse from the loop regions. While we find spectral index differences 
between loop-like filaments in the interior and exterior regions of the 
nebula, the void regions have spectra 
similar to those of the corresponding filaments (Table 1), although the larger
uncertainties on these values prevent us from ruling out softer spectra
from regions outside of the filaments, as might be expected if particles are
diffusing away from regions of acceleration. The jet-like feature
extending westward from the pulsar appears to merge with a loop structure
to the southwest, possibly suggesting that the former is not actually 
a jet. However, the spectrum of this feature is flatter than
for the adjacent loop, indicating that the jet is a distinct feature.

\begin{table*}
\begin{center}
\caption{Spectral Indices: Jet, Loop, and Void Regions}
\begin{tabular}{cccccccccccc}\\ \hline \hline
\multicolumn{5}{c}{Jet} & \multicolumn{4}{c}{Loops} & \multicolumn{3}{c}
{Voids} \\ \hline
\vspace{1mm}
Whole & A & B & C &~~~ & A & B & C &~~~& A & B & C \\
2.06 &1.93 & 2.14  & 2.05 &&
2.29 & 2.14 & 2.17 &&
2.27  & 2.27 & 2.31 \\
$\pm 0.04$ & $\pm 0.11$ & $\pm 0.07$ & $\pm 0.07$ &&
$\pm 0.04$ & $\pm 0.06$ & $\pm 0.08$ &&
$\pm 0.07$ & $\pm 0.11$ & $\pm 0.11$ \\ \hline
\end{tabular}
\end{center}
\vspace{-0.5cm}
\end{table*}

The overall morphology of 3C~58 reflects a symmetry associated with the pulsar
rotation axis, whose projection is believed to lie in the E-W direction 
(Slane, Helfand, \& Murray 2002). Blandford (2002) has suggested that the 
current flow in such pulsar-driven nebulae is directed along the polar 
regions, with a return
current in the equatorial plane (or vice versa). This suggests possible 
spectral differences between the polar/equatorial regions and those regions
in between. To investigate this, we extracted spectra in cone-like regions
surrounding the pulsar (see Figure 2). We again restricted the energy band 
to $2.2 - 8$~keV and performed fits to a power law model. As shown in 
Figure 5, although there is some variation in the average index values,
we find no obvious azimuthal variation in the spectral index that 
might be associated with this suggested large-scale current flow.

We note that the radial variation in spectral index shown in Figure 4 also
appears to be at odds with a picture in which large-scale current 
flows connected
with the pulsar extend to the outer reaches of the nebula. Rather, the trend
appears more consistent with diffusion of particles injected into the central
regions of the PWN. Indeed, the surface brightness of the nebula decreases
with radius (Figure 6), and the decline is most dramatic for high 
energies, as expected for synchrotron aging. However, we note that the 
radial profile in the 0.5--1.0~keV band shows a bump at large radii, suggestive
of a shell of soft emission.

The X-ray spectrum from the outermost region of 3C~58 (``Shell'' region in
Figure 2)\footnote{
Note that ``Shell'' region has been chosen to encompass emission only from
the outer portions of 3C~58 that fall on the S3 chip because the 
adjacent S2 chip has considerably reduced low energy response, and
thus is not sensitive to the soft thermal emission. 
}
is not well fit by a simple power law. Residual emission 
below $\sim 1$~keV requires an additional thermal model with $kT \sim 
0.25$~keV and roughly a factor of 3 overabundance of Ne. This is similar 
to the results reported by Bocchino et al. (2001) based on {\em XMM-Newton} 
observations, and represents confirmation of the long-sought thermal shell of
3C~58. (See $\S 4.6$ and Figure 9 for further details.)

\section{Discussion}

\subsection{Large-Scale Structure of PWN}

The large-scale elongated shape of 3C~58 is similar to that found (particularly
in the radio band) for a number of other PWNe including the Crab Nebula and 
G54.1+0.3. Magnetohydrodynamical calculations by Begelman \& Li (1992) 
and van der Swaluw (2003) show that such an elongation can result from the 
pinching effect of a toroidal magnetic field for which the projected axis
lies along the 
long axis of the PWN. The pinching effect results in a low pressure at the 
edge of the bubble along the major axis with respect to the (higher) pressure 
at the edge of the minor axis, which yields the elongated structure.
The elongation thus marks the projection of the spin 
axis of the pulsar producing the wound-up field.  In 3C~58 this is
consistent with the inference of an E-W direction for the projected 
spin axis based on the interpretation of the extended structure in the inner 
nebula as being associated with a tilted ring-shaped wind termination shock
zone (Slane, Helfand, \& Murray 2002 -- see, also, Section 4.2). 

\begin{figure}[t]
\plotone{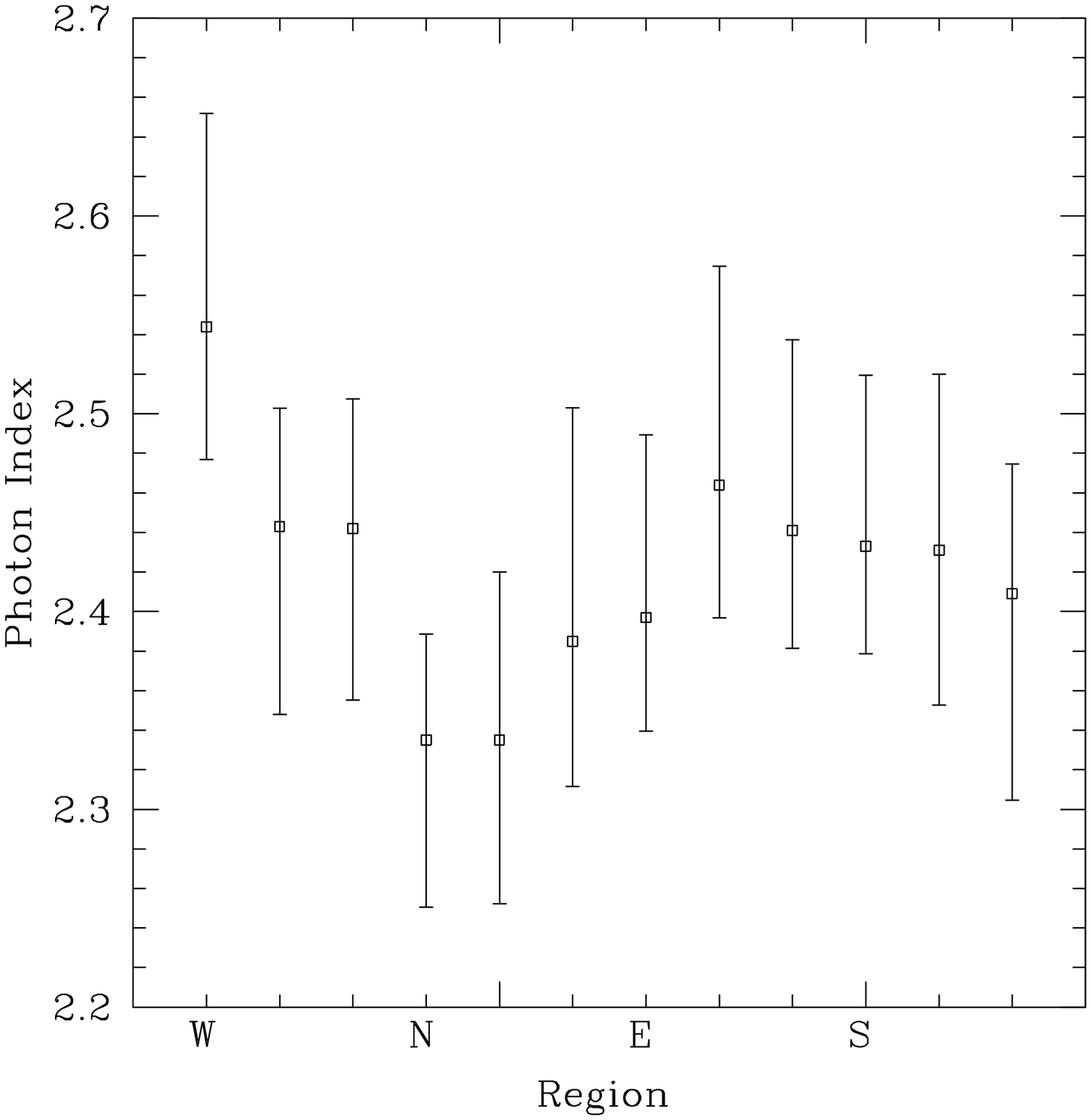}
\caption{
Spectral index as a function of azimuth in 3C~58. Here spectra were
extracted in cone-like regions extending from, but not including,
the pulsar and its surrounding toroid.
}
\end{figure}

Radio polarization measurements at 6, 21, and 50 cm show a complex
magnetic field in the central regions of 3C~58 (Wilson \& Weiler 1976),
with some indications of a toroidal morphology. In the outer regions of the
nebula, however, the component of the nebular magnetic field in the plane 
of the sky is predominantly oriented in an E-W direction, possibly indicating
that the toroidal structure has been disturbed. 
However, small-scale variations in interstellar Faraday rotation could not
not be ruled out as a source of the inferred polarization structure, and
higher resolution measurements are clearly of interest.
Begelman (1998) has noted
that toroidal fields are susceptible to kink instabilities that may
destroy the toroidal geometry. This could explain the lack of larger scale
polarization evidence for a toroidal field. 
We discuss further evidence for the
disruption of the toroidal field in Section 4.5.

\begin{figure}[t]
\plotone{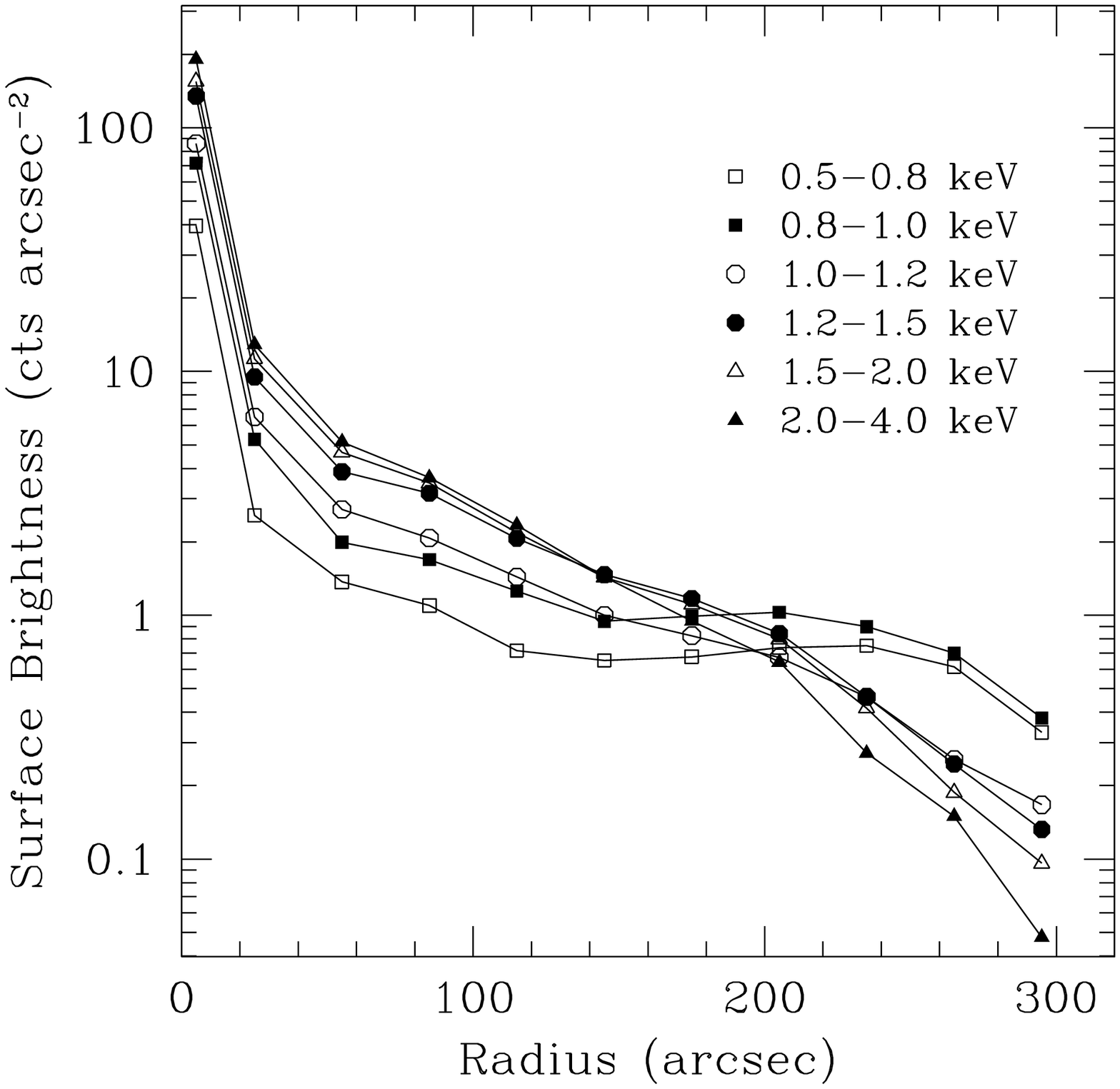}
\caption{
Surface brightness as a function of energy for the southwest
sector in 3C~58.
}
\end{figure}

The magnetic field strength inferred for 3C~58 from the minimum energy condition
based on the radio spectrum is $\sim 80 \mu$G (Green \& Scheuer 1992). Using 
the historical age, this field results in a synchrotron burn-off break at
$\nu_{br} \approx 3 \times 10^{15}$~Hz, or an energy of $\sim 0.01$~keV.
This is consistent with fact that the nebula has very nearly the same size
in the radio and X-ray bands; the effects of synchrotron aging become evident
only just below the soft X-ray band (and are indeed observed in that band, as
seen in Fig. 3), and only the electrons corresponding to the higher energy
X-rays are appreciably diminished far from the pulsar. 
In fact, MHD calculations (Kennel \& Coroniti 1984; hereafter
KC84) show
that the magnetic field strength initially increases with radius and
reaches the equipartition value at a radius of $\sim 3-10 r_s,$ depending
upon the value of the magnetization parameter
$$ \sigma \equiv \frac{B^2}{4 \pi n \gamma m c^2}$$
which represents the ratio of the Poynting flux to particle flux in the wind.
Here $n$ is the density, $\gamma$ is the Lorentz factor of the wind, and 
$r_s$ is the termination shock radius defined by balancing the ram
pressure of the wind with the internal pressure in the nebula, $P_n$:
$$ r_s^2 = \frac{\dot E}{4 \pi \xi c P_n},$$
where $\dot E$ is the spin-down power of the pulsar and $\xi$ is the fraction
of a sphere covered by the wind. Beyond this point the field decreases
as $\sim 1/r$ which means the effective synchrotron break extends to even 
higher energies.

Using the solutions of KC84, Reynolds (2003) has calculated the expected
variation in spectral index in the soft X-ray band given synchrotron losses 
in the nebula. For any reasonable values of $\sigma,$ the expected photon
index
increases much more rapidly with radius in the outer nebula than we
observe in 3C~58 (Fig. 3). Such a deviation from the predicted profile
is also observed for G21.5--0.9 (Slane et al. 2000). This may be additional
evidence for a magnetic field structure that deviates from the pure toroidal
structure assumed by KC84, supporting the suggestion that kink instabilities
disrupt the internal field structure (Begelman 1998).

We note that 3C~58 is not quite symmetric about the position of its pulsar,
which is located closer to the western edge of the nebula. This may be an
indication of a slightly lower density in the eastern direction, or could be
associated with motion of the pulsar itself. For a simple geometric assumption
based on the offset between the PWN center and the pulsar position, the latter 
interpretation would require a velocity of $\sim 500 {\rm\ km\ s}^{-1}$ in 
the plane of the sky, which is fast but not unreasonable for a pulsar. 
However, when the hydrodynamics are considered this scenario appears
problematic. The high sound speed in the PWN quickly smoothes out any pressure
differences in the nebula, resulting in symmetric expansion 
(van der Swaluw, Downes \& Keegan 2003). It is possible that the effects of
internal magnetic fields alter this picture considerably; 3-D MHD simulations
are needed to probe this more fully.

\begin{figure}[t]
\plotone{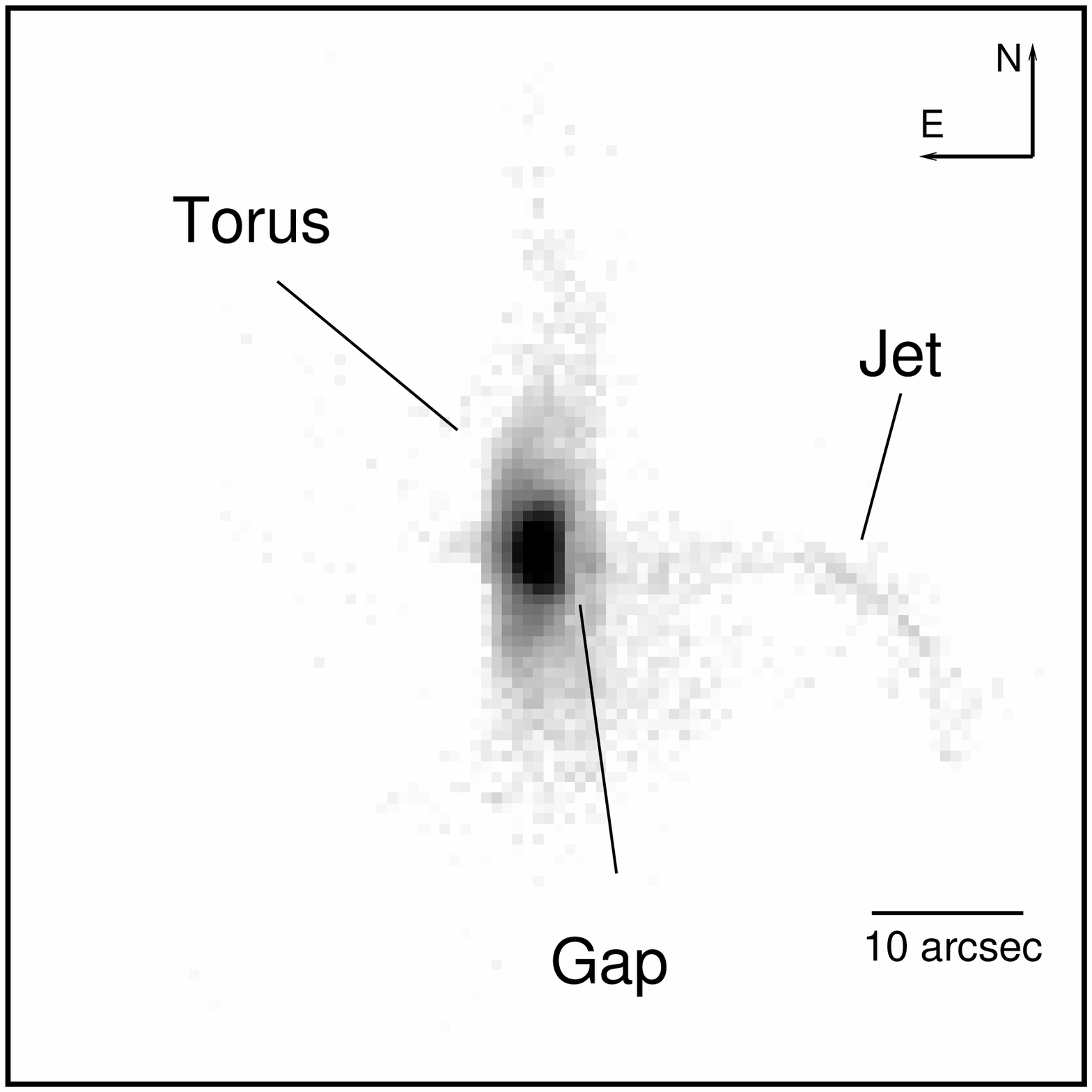}
\caption{
\chandra\ image of the innermost region of 3C~58 showing the neutron star
embedded in an elongated structure. A gap in the emission profile to the
west of the pulsar (see Fig. 5) suggests a ring/torus structure. A jet extends
to the west, with a hint of a counterjet component in the east.
}
\end{figure}

An alternative scenario would be that the western portion of the
PWN has begun interacting with the reverse shock formed when the expanding
ejecta meets the swept-up shell from the supernova blast wave. 
This could result if the ambient density is greater in the western
direction, or if the pulsar is moving in that direction. In either case
the compression of the PWN along this side naturally
leads to an asymmetry of the PWN structure (see for example
Figure 5 in van der Swaluw, Downes \& Keegan 2003).
The reverse shock scenario has been suggested by several authors
as an explanation of the velocity/age discrepancy in \snr\ (van der Swaluw
2003) and its low-frequency spectral break (Gallant et al. 2002).
A serious drawback for this scenario is that in order for the reverse shock 
to propagate back to a distance of $\sim 4$~pc from the 
PWN center (i.e., to the observed western edge of the PWN), the required 
density of the medium surrounding the complete SNR is sufficiently high
that one would expect to observe significant emission from an extended SNR 
shell, which is not the case.
An estimate for this density can be obtained using the expansion
model of a SNR from Truelove \& McKee (1999), which yields roughly
$n_0\sim 0.21$ under the assumption that the energy of the SNR
equals $10^{51}$ ergs. For lower explosion energies the inferred density
will decrease.

\begin{figure}[t]
\plotone{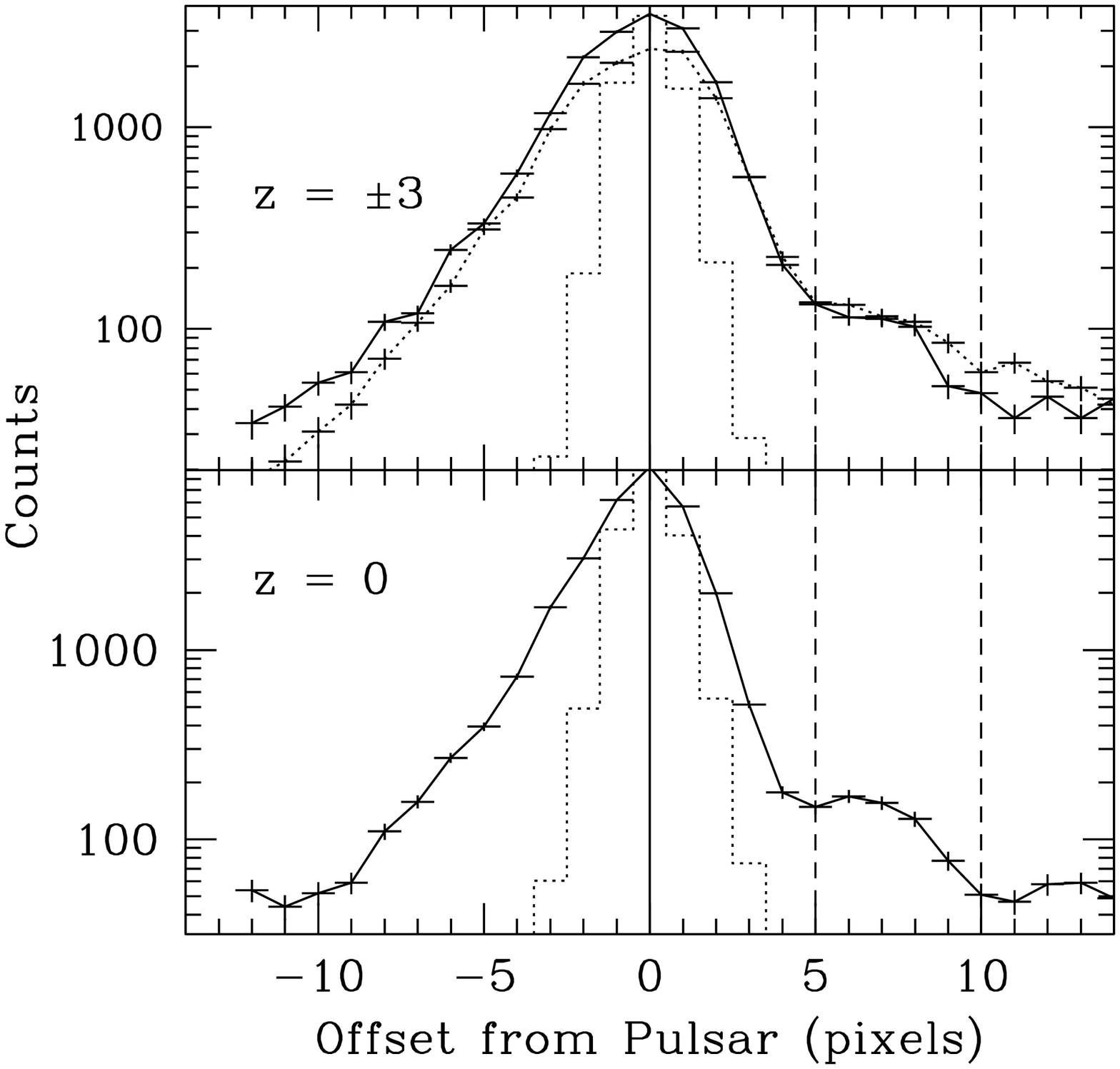}
\caption{
Brightness profiles extracted from horizontal slices through the inner
nebula of 3C~58 (Fig. 4). The lower curve is from a three-pixel-wide box
centered on the pulsar. The upper curve is from identical boxes centered
three pixels above (solid) and below (dashed) the pulsar position. The
histogram in each plot corresponds to the \chandra/ACIS profile for a
point source. (Each pixel is $\sim 0.49$ arcsec wide.)
}
\end{figure}

\subsection{The Inner Nebula}

The innermost region of 3C~58 consists of a bright, elongated compact 
structure centered on the pulsar J0205+6449. This inner nebulosity 
is bounded along the western edge by a radio wisp (Frail \& Moffet 1993)
and was interpreted by Slane, Helfand, \& Murray (2002) as emission
from the pulsar wind termination shock. Our deeper observation modifies
this interpretation slightly. As illustrated in Figure 7, the image of the 
central region is suggestive of a toroidal structure that is tilted about 
a north-south axis, with the pulsar at its center, and with a jet-like 
feature extending to the west. In Figure 8 we plot brightness profiles in 
three-pixel-wide slices through the nebula (in the east-west direction). In the 
lower panel, the profile extends through the pulsar position, while the upper 
panel shows profiles centered three pixels ($1.5^{\prime\prime}$) north and south of the pulsar. The profile of the central emission is 
extended well beyond the $\sim 1^{\prime\prime}$ point-spread function of 
the telescope, indicating that diffuse emission surrounds the pulsar.
Immediately evident is a bump in the profile centered $\sim 
3.5^{\prime\prime}$ to the west of the pulsar. We interpret this as
evidence of a ring surrounding the core of the nebula. Presumably, much 
like in the Crab Nebula (Weisskopf et al. 2000), we are seeing evidence
of an inner diffuse emission region associated with the termination shock, and
a surrounding torus. The brightness profile is asymmetric about the
pulsar position, falling off more rapidly along the western side. This
may indicate that the eastern side is beamed toward us, implying that the
torus is tilted out of the plane of the sky on this side. 

\begin{table*}
\begin{center}
\caption{Spectrum of Neutron Star and Torus}
\begin{tabular}{ccccccc}\\ \hline \hline
Model$^a$ & $N_H^b$ & $\Gamma$ & $F_x^c$ & $T_\infty$(MK)
& $R_\infty$(km) & $\chi^2$/dof \\ \hline
\\
\multicolumn{7}{c}{Neutron Star} \\
PL & $3.4 \pm 0.1$ & $1.67 \pm 0.03$ & 0.91
& N/A & N/A & 431.5/304 \\
PL + BB & $6.8 \pm 0.8$ & $1.78^{+.02}_{-.04}$ & 1.02
& $1.3 \pm 0.1$  & 10.7 & 357.5/302 \\
PL + NSA & $4.7 \pm 0.2$ & $1.65 \pm 0.03$ & 0.92
& $1.08 \pm 0.03$  & 10.0 & 398.6/303 \\
\\
\multicolumn{7}{c}{Torus} \\
PL & $4.2 \pm 0.3$ & $1.87 \pm 0.02$ & 4.06
& N/A & N/A & 578.2/418 \\ \hline
\multicolumn{7}{l}{(a) Models are: power law (PL), blackbody (BB), and
neutron star atmosphere (NSA)} \\
\multicolumn{7}{l}{(b) in units of $10^{21}{\rm\ cm}^{-2}$} \\
\multicolumn{7}{l}{(c) Unabsorbed $0.5 - 10$~keV flux (in units of
$10^{-12}$~erg~cm$^{-2}$~s$^{-1}$)}
\end{tabular}
\end{center}
\vspace{-0.5cm}
\end{table*}

The gap between the innermost extended region and the apparent ring feature
to the west is located $\sim 2.5$~arcsec from the pulsar. A similar feature,
though marginally significant, is observed northward of the pulsar at a
distance of $\sim 8$~arcsec. If interpreted as a circular termination shock
zone, the inferred inclination angle in the plane of the sky is roughly
70 degrees, similar to that inferred from the outer radii of the toroidal
structure (Slane, Helfand, \& Murray 2002).
The luminosity of the toroidal region is $L_x (0.5 - 10 {\rm\ keV})
= 5.3 \times 10^{33} {\rm\ erg\ s}^{-1}$.

\subsection{The Neutron Star}
Given the young age of 3C~58, the X-ray emission from \psr\ places strong 
constraints on the rate at which the neutron star interior has cooled. Using
a $\sim 35$~ks ACIS observation of \snr, Slane, Helfand, \& Murray (2002)
derived an upper limit of $T < 1.08 \times 10^6$~K for the surface temperature 
of the neutron star. 
Following the same technique for  our deeper observation described here,
we have extracted the spectrum from the $3 \times 3$ 
pixel region centered on the pulsar; spectral fit results are summarized
in Table 2. With the increased sensitivity of this
exposure, we find that a power law model no longer adequately describes the
spectrum. The best-fit model yields a reduced chi-squared value of
$\chi_r^2 = 1.4.$ There are significant residuals around the Si-K edge that
suggest calibration-related effects, but the model also requires a 
column density significantly lower than that derived above for the nebula. 
Addition of a blackbody component improves the fit significantly
($\chi^2_r = 1.2$), but requires a column density
larger than that for the PWN. With $N_H$ constrained to the PWN value, 
the blackbody component requires an emitting radius of only $\sim 2.6$~km 
and a temperature of $kT = 0.15$~keV.  Constraining the size of the emitting
region to that for a neutron star with $R_\infty = 12$~km yields a $3 \sigma$
upper limit of 0.086~keV ($T < 1.02 \times 10^6$~K) 
on the temperature for emission from the entire surface based on fixing the
column density at the upper limit of the 90\% confidence level for that of
the PWN.

Using a model for neutron star emission with a hydrogen atmosphere 
(Pavlov et al. 1995) and a surface magnetic field of $B = 10^{12}$~G
yields a significant improvement to the power law fit as well 
($\chi_r^2 = 1.3$). Here we have fixed the NS mass at 1.4~$M_\odot$,
the radius at 10~km, and the distance at 3.2~kpc. The best-fit models
yields $T = (1.08 \pm 0.3) \times 10^6$~K. 

We note that the emission from \psr\ is not clearly resolved from the 
contributions of the surrounding compact nebula, and the spectral index of 
the pulsar may differ from that which we derived for the central emission. 
RXTE observations (Ransom et al. 2004) indicate a power law photon index
of 1.0 for the pulsed component of the emission from \psr. The pulsed fraction
derived from \chandra\ HRC observations (Murray et al. 2002) 
is $\sim 20\%$ in the
$0.08 - 10$~keV band. Accounting for the ratio of encircled energy fractions 
for the HRC and ACIS source extraction regions, and also for the relative 
effective areas for the detectors, the expected ACIS count rate from this
component is $1.6 \times 10^{-2} {\rm\ cts\ s}^{-1}$. Adding a component
with these fixed parameters to the spectral model improves the spectral
fit. Thus, while we do not uniquely identify such a spectral component, our
results are consistent with those derived from RXTE observations. Moreover,
if the neutron star atmosphere model described above is added to the model 
as well, the fit is improved again (with an F-test probability of
$3 \times 10^{-6}$ that the improvement is random) and yields a NS temperature
of $T = (1.00 \pm 0.04) \times 10^6$~K and a column density
$N_H = (4.8 \pm 0.2) \times 10^{21}{\rm\ cm}^{-2}$ that is consistent
with that for the PWN as a whole.

This temperature (as well as that derived without the second power law 
component, and the  upper limit for blackbody emission from 
the entire surface) falls far below
the predictions for standard cooling by the modified Urca process for a
$1.4 M_\odot$ neutron star with a radius of 12~km. For larger masses, it
is possible for the proton fraction in the interior to become sufficiently
high to permit cooling through the direct Urca process (Kaminker,
Yakovlev, \&  Gnedin 2002; Yakovlev et al. 2002), thus enabling the upper limit 
for \psr\ to be 
accommodated. Alternatively, the presence of pion condensates (or other
exotic particles) in the core would also lead to rapid cooling that results
in such low temperatures (Tsuruta et al. 2002). 
The results presented here, combined with those for the Vela Pulsar (Pavlov
et al. 2001) and other young neutron stars point to rapid cooling
by nonstandard processes (i.e. something other than the modified Urca
process). Page et al. (2004) identify ``minimal cooling'' scenarios which
are able to accommodate the results for pulsars other than that in \snr.
Recent limits for the compact object in CTA~1 (Slane et al. 2004; Halpern
et al. 2004), as well as limits for undetected neutron stars in several
nearby SNRs (Kaplan et al. 2004) may provide additional evidence for such
``enhanced cooling'' in neutron stars. Further observations are of 
considerable importance in constraining cooling models.

Thus, while the results
reported here, combined with those for the Vela Pulsar and
a number of other young neutron stars, point rather clearly to 
rapid cooling by non-standard processing (i.e., something other than the
modified Urca process), further observations of other young neutron
stars are required to differentiate
between competing theoretical models.

\subsection{The Jet-Like Feature}
The elongated structure extending westward from the position of the pulsar has
the appearance of a jet. Its orientation is consistent, in projection, with the 
pulsar rotation axis inferred from the wind termination shock region discussed 
above, and also the east-west elongation of the entire PWN. The structure shows
considerable curvature, similar to that seen in the Crab Pulsar jet. 
The power law spectrum of the jet-like structure in 3C~58 has a photon 
index of $2.06 \pm 0.04$.  The spectral index does not vary
along the length of the feature. The observed luminosity is
$L_x (0.5 - 10 {\rm\ keV}) = 6.8 \times 10^{32} d_{3.2}^2 {\rm\ erg\ s}^{-1}$,
nearly a factor of 10 smaller than that for the torus. 
For the Crab Nebula, the torus is nearly 20 times more luminous than the
jet in X-rays (F. Seward, private communication), while for PSR~B1509--58 the
jet is brighter than the extended inner emission (Gaensler et al. 2002).

Approximating the structure as a cylinder whose length is 35~arcsec, with
a width of 6~arcsec, the volume is $V \approx 1.2 \times 10^{53} \phi 
d_{3.2}^3 {\rm\ cm}^3$ where $\phi$ is the volume filling factor. 
Using the X-ray spectrum, assuming synchrotron emission, the
minimum total energy in this volume is then
$E_{min} = 1.7 \times 10^{44} (1 + k)^{4/7} \phi^{3/7} d_{3.2}^{17/7}
{\rm\ erg}$ (see, e.g., Gaensler et al. 2001); here $k$ is the ratio of 
ion to electron energy. The corresponding 
magnetic field $B_{min} = 3.5 \times 10^{-5} (1 + k)^{2/7} \phi^{-2/7} 
d_{3.2}^{-2/7} {\rm\ G}$. We note that this is smaller than the equipartition
field for the nebula as a whole, $B_{eq} \approx 80 \mu$G as determined from 
radio measurements (Green \& Scheuer 1992). This is consistent with the
expectation that the magnetic field is lower in the central regions of
the PWN, and rises to the equipartition value only at larger radii (KC84).

Since there is no spectral variation across the elongated structure, the
synchrotron lifetime of the energetic particles constrains the rate at which 
energy is being injected. For particles radiating X-rays with energies
of 5~keV, the synchrotron lifetime is $t_{s} = 4.4 \times 10^9 (1 + k)^{-3/7} 
\phi^{3/7} d_{3.2}^{3/7} {\rm\ s}$. The outflow velocity of the material
must then be $v_j > l/t_s = 3.4 \times 10^8 (1 + k)^{3/7} \phi^{-3/7} 
d_{3.2}^{4/7} {\rm\ cm\ s}^{-1}$, or $v_j \ga 0.01 c$. Note that for the
Crab, $v_j \la 0.03 c$ (Willingale et al. 2001), a similar value.

For the X-ray emitting particles in the jet-like structure, the rate at
which energy must be supplied by the pulsar is $\dot E_{j} \geq E_{min}/t_s
= 8.3 \times 10^{33} (1 + k) d_{3.2}^{2} {\rm\ cm\ s}^{-1}$. This represents
only $0.03\%$ of the energy loss rate from the pulsar itself
($\dot E = 2.7 \times 10^{37}{\rm\ ergs\ s}^{-1}$). The value
for the jet in PSR~B1509--58 is 0.5\%.
indicating that the 3C~58 jet (if it is indeed a jet) carries proportionally 
much less of the pulsar spin-down power in X-rays. 

\subsection{Filamentary Loop Structures}

Our deep \chandra\ observation of \snr\ reveals a complex of loop-like
filaments most prominent near the central regions (Figure 1), but evident 
throughout the nebula. These structures, whose X-ray spectra are nonthermal,
are very well correlated with features observed in the radio band (Figure 2).
Extensive filamentary structure is also observed in radio images of the Crab
Nebula. In this case, the filaments coincide with optical filaments observed
in H$\alpha$, [OIII], and other lines, indicating thermal emission. MHD
simulations suggest that the Crab filaments form from Rayleigh-Taylor 
instabilities as the expanding relativistic bubble encounters slower moving ejecta 
(Jun 1998; Bucciantini 2004). The enhanced radio emission presumably results 
from compression of the relativistic fluid in the expanding nebula as it 
encounters these embedded filaments. If this is correct, one might also expect 
X-ray emission from these radio filaments. In the Crab Nebula, however, the 
magnetic field is sufficiently high that the X-ray emitting electrons do 
not reach the outer portions of the nebula as a consequence of synchrotron
losses. Because the radio filaments appear to be concentrated in a shell,
there is very little X-ray emission coincident with these structures;
indeed, the X-ray emission region is much smaller in extent than the
radio nebula.\footnote{
Deep \chandra\ images of the Crab Nebula do reveal several filaments outside
the bright X-ray torus that coincide with radio filaments (F. Seward - private
communication).}

Optical observations of \snr\ reveal faint thermal filaments as well (van den 
Bergh 1978), which presumably have a similar origin. The velocities of these
filaments are $\sim \pm 900{\rm\ km\ s}^{-1}$ (Fesen 1983), sufficiently
high to indicate that the PWN is young, but too small to account for the
current size of \snr\ if the historical age is assumed -- one of 
several standing problems with regard to its evolution.
In Figure 2 we present an H$\alpha$ image obtained in a 8400~s integration 
with the 1.3~m McGraw-Hill telescope, along with a single contour marking the
outer boundary of the radio nebula. The interior is filled with filaments, 
some organized into apparent ring-like shapes. Of particular note is the 
$\sim 40$~arcsec ring in the southwestern portion of the nebula,
which coincides with an identical feature in the X-ray image, as
well as in the radio image. This feature appears to be an example of the
filamentary structures described above. However, it is apparent from Figure
2 that many of the X-ray filaments do not have corresponding optical 
structures. While comparisons with deeper optical images are clearly
needed, the fact that many of the X-ray features without optical counterparts
are brighter than average in X-rays suggests that these may actually
arise from a different mechanism.

We propose that the bulk of the discrete structures seen in the X-ray 
and radio images of 3C~58 are magnetic loops torn from the toroidal field 
by kink instabilities. In the inner nebula, the loop sizes are similar to 
the size of the termination shock radius, as suggested by Begelman (1998). 
As the structures expand, they enlarge slightly as a consequence of the
decreasing pressure 
in the nebula. Some of the observed X-ray structure in the outermost regions
may be the result of thermal filaments produced by Rayleigh-Taylor 
instabilities, similar to the filaments in the Crab Nebula. The observed 
soft X-ray shell (see Section 4.6) demonstrates the presence of ejecta in 
these outer regions, and the optical/radio/X-ray emission from the southwest 
ring appears consistent with this scenario. While some of the optical 
filaments do appear to be located in the central regions, these may lie 
primarily along a shell seen in projection.

We note that considerable loop-like filamentary structure is evident in
\chandra\ observations of the Crab Nebula (Weisskopf et al. 2000). These 
features are primarily observed encircling the bright Crab torus, perpendicular
to the toroidal plane, and may result from currents within the torus itself. 
It is at least conceivable that such currents are signatures of the kink 
instabilities suggested above.

If the X-ray filaments in the inner portions of \snr\ are indeed the result of a
disrupted toroidal magnetic field, this may also explain the radio polarization
results that show a primarily radial field in the outer regions (Wilson
\& Weiler 1976) as well as the discrepancy between the predicted and
observed radial variation in the X-ray spectral index (Reynolds 2003).
The large-scale structure of the nebula, which is well-described by MHD
models invoking pinching by a toroidal field to confine the equatorial
region (Begelman \& Li 1992; van der Swaluw 2003), would seem to argue 
against this. However, it is possible that 
the innate asymmetry that results in this large-scale morphology is
imprinted in the flow close to the termination shock, and that disruption
of the field beyond this point does not strongly affect the expansion.
New MHD modeling is necessary to investigate more fully the self-consistency
of this scenario.

\begin{figure}[t]
\plotone{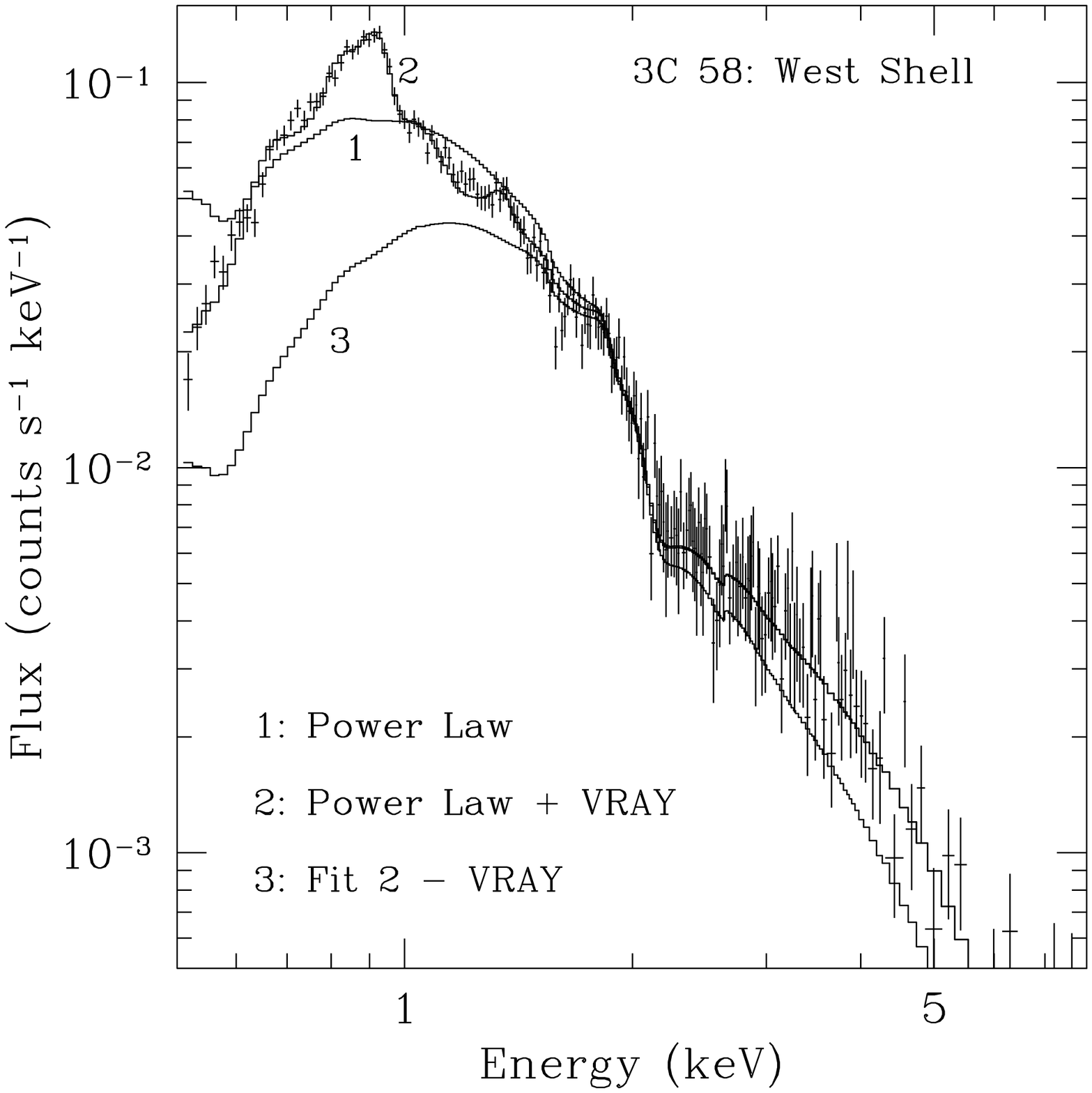}
\caption{
ACIS-S spectrum of the outer region of 3C~58 (see text for
description). Histograms show the best fit power law model(1), the best
fit power law + equilibrium plasma model (2), and the power law component
only from fit 2 (3).
}
\end{figure}

\subsection{The Thermal Shell}

Studies of 3C~58 with {\em XMM-Newton} reveal evidence for thermal emission
in the outer regions of the nebula (Bocchino et al. 2001). Our observations
confirm this result and provide improved parameters describing the nature
of the emission. As shown in Figure 6, the X-ray surface brightness for \snr\ 
declines monotonically with radius except at energies below $\sim 1$~keV
where there is a noticeable plateau in the outer regions, suggestive of a
shell of soft emission. The spectrum from the outer shell of the nebula is
shown in Figure 9. Contrary to what is found for all regions in the nebula
interior, a single power law model is not able to adequately fit the spectrum
of the outer region. The best-fit power law is indicated in the Figure as the
solid curve labeled 1; there is significant residual emission between $\sim 0.7
- 1.0$~keV. Adding a single equilibrium plasma model (Raymond \& Smith
1977) with
$kT \sim 0.25$~keV and enhanced Ne and Mg abundances provides an excellent 
fit to the data (curve 2). The curve labeled 3 in the Figure
corresponds to power law component only for the best fit two-component model,
illustrating the reasonably large contribution from the thermal shell.

Similar results are obtained with the {\tt vpshock} model (Borkowski,
Lyerly, \& Reynolds 2001). The equilibrium model provides a slightly better
fit, and the vpshock model requires an ionization timescale appropriate
for equilibrium.  The best-fit values for the thermal shell are 
given in Table 3. 

\pspicture(0,-0.9)(8.5,4)
\rput[tl]{0}(-0.5,3.8){
\begin{minipage}{8.25cm}
\small\parindent=3.5mm
\begin{center}
TABLE 3

{\sc Spectrum of Thermal Shell}

\vspace{1mm}

\begin{tabular}{cc}\\ \hline\hline
Parameter & Value \\ \hline
$N_H$ & $4.5 \times 10^{21} {\rm\ cm}^{-2}$ (fixed) \\
$kT$ & $0.23 \pm 0.01$ keV \\
$K^{(\rm a)}$ & $6.2 \times 10^{-4}$ \\
$[Ne]/[Ne]_\odot$ & $3.1 \pm 0.2$ \\
$[Mg]/[Mg]_\odot$ & $2.0 \pm 0.6$ \\
$\Gamma$ & $2.74 \pm 0.05$ \\ \hline
(a) $K = [10^{-14}/(4 \pi d^2)] \int n_e n_H dV$
\end{tabular}
\end{center}
\end{minipage}
}
\endpspicture

Treating \snr\ as a prolate ellipsoid, we estimate
a total volume of $\sim 3 \times 10^{57} d_{3.2}^3 {\rm\ cm}^3$. The region
from which the shell spectrum has been extracted is complicated; we estimate
its volume as roughly one-third that of the entire PWN, but this quantity
is quite uncertain. Based on this estimate, the volume emission measure
from the spectral fit implies a density of $n \sim 0.3 d_{3.2}^{-1/2}
{\rm\ cm}^{-3}$ for the thermal material. This corresponds to a mass
of $\sim 0.5 d_{3.2}^{5/2} M_\odot$ in this volume, and roughly two 
times this amount for the entire shell (since less than half of the
shell volume is contained in the spectral extraction region). Spectral fits
for the inner regions of 3C~58 reveal emission from this same plasma component;
this mass estimate is thus a lower limit.

The enhanced abundance of Ne and Mg indicates that the shell is composed 
of ejecta.
This is consistent with the picture wherein the optical filaments in \snr\ 
result from interactions between the expanding bubble and the supernova
ejecta. However, the mass estimated above appears to be in conflict with the
assumed age for \snr. Chevalier (2003) notes that the amount of mass
swept up by the PWN is $M_{sw} \approx \dot E R^{-2} t^3$ for most 
supernova density distributions; here $R$ is the radius of the PWN. For
\snr\ this leads to a predicted swept-up mass nearly 500 times smaller
than that inferred from the spectral fits. A more detailed study of this
thermal component is currently underway using a large {\em XMM-Newton}
dataset obtained from numerous calibration observations, and we defer
further discussion of this emission component and its implications for
the dynamics and evolution of 3C~58 to a future publication.

\section{Summary}

Deep \chandra\ observations of \snr\ reveal a rich X-ray structure consisting
of what appear to be magnetic loops. These structures predominate in the
interior, just outside the pulsar termination shock, but are also found
in outer regions. The structures are well correlated with radio features,
but not, in general, with the optical filaments in \snr. 
They appear to have a different
origin from the optical and radio filaments in the Crab Nebula, possibly
originating from kink instabilities in the inner toroidal field. This
notion may find additional support from radio polarization data, which 
appear to show a significant change in the field morphology away from
the central regions, but improved radio observations are required to confirm
this. The observed radial variation in the X-ray spectral index is shallower
than expected from the KC84 model, possibly also indicating that
the field structure differs from that assumed, although the discrepancy could
arise from factors other than a modification of the toroidal field.

The detection of a thermal shell confirms the earlier reports by Bocchino
et al. (2001) based on XMM data. The emission is enriched in Ne and Mg,
indicating that it is swept-up ejecta from the progenitor. The total mass
of X-ray emitting ejecta is relatively uncertain due to the poorly defined
volume and filling factor, as well as the dominant contribution from 
synchrotron emission throughout most of the PWN. It is at least several
$0.5 - 1 M_\odot$.

The X-ray spectrum of \psr\ in \snr\ is not well described by a single
power law. Evidence consistent with a hard spectrum ($\Gamma = 1$) associated
with the pulsed component observed with the HRC and RXTE is revealed,
as is the apparent detection of thermal emission from a hydrogen atmosphere
of a neutron star. At the most conservative level, new limits on any
blackbody-like emission imply a surface temperature $T < 1.02 \times 10^6$~K
which is far below predictions of standard models based on neutrino cooling
through the modified Urca process, forcing one to consider either direct Urca 
cooling due to an increased proton fraction or the presence of exotic particles
in the interior.

Immediately surrounding the pulsar, the X-ray emission reveals an elongated
structure that appears to be a tilted toroid along with a jet extended to the
west and some evidence of a counterjet. In this regard, the inner portions of
3C~58 appear similar to the Crab nebula -- if not in brightness, then in 
overall structure. The inferred position of the termination shock is inside
the position of the radio wisp discovered by Frail and Moffett (1993), 
presumably indicating that this wisp bounds the outer torus. The X-ray
spectrum of the jet shows no evidence of synchrotron aging, allowing us to
to infer a flow velocity $v_j \ga 0.01 c$, similar to that for the Crab jet.

The large-scale morphology of \snr, with its elongation in a
direction perpendicular to the inferred pulsar rotation axis, is consistent
with models for PWNe with embedded toroidal magnetic fields, but the 
radial variation in spectral index and the radio polarization observations
seem to indicate that the field is more complicated than this, perhaps
consistent with the small-scale loops described above. More sensitive radio
polarization measurements are required to investigate this further, and
deeper optical observations are also of interest to investigate more fully the 
relationship between optical filaments and the observed X-ray and radio 
structures.


\acknowledgments
This work was supported in part by the National Aeronautics and
Space Administration through contract NAS8-39073 and grant GO0-1117A (POS).
We thank Ms. Eve Armstrong and Ms. Sarah Tuttle for obtaining the optical
observations of 3C58, and Mr. Nestor Mirabel for reducing the data.
POS also acknowledges helpful discussions with Roger Chevalier and Bryan
Gaensler. Portions of this work were carried out in conjunction with the
``Physics of SNRs in the Chandra, XMM-Newton, and INTEGRAL Era'' workshop
held at the International Space Science Institute, Bern, Switzerland.





\end{document}